# The analytical solution to the migration of an epithelial monolayer with a circular spreading front and its implications in the gap closure process

Tiankai Zhao[*], Hongyan Yuan

Shenzhen Key Laboratory of Soft Mechanics & Smart Manufacturing, Department of Mechanics and Aerospace Engineering, Southern University of Science and Technology, Shenzhen, 518055, China

*Corresponding Author:　zhaotk@sustech.edu.cn

**Abstract**

The coordinated behaviors of epithelial cells are widely observed in tissue development, such as re-epithelialization, tumor growth, and morphogenesis. In these processes, cells either migrate collectively or organize themselves into specific structures to serve certain purposes. In this work, we study a spreading epithelial monolayer whose migrating front encloses a circular gap in the monolayer center. Such tissue is usually used to mimic the wound healing process in Virto. We model the epithelial sheet as a layer of active viscous polar fluid. With an axisymmetric assumption, the model can be analytically solved under two special conditions, suggesting two possible spreading modes for the epithelial monolayer. Based on these two sets of analytical solutions, we assess the velocity of the spreading front affected by the gap size, the active intercellular contractility, and the purse-string contraction acting on the spreading edge. Several critical values exist in the model parameters for the initiation of the gap closure process, and the purse-string contraction plays a vital role in governing the gap closure kinetics. Finally, the instability of the morphology of the spreading front was studied. Numerical calculations show how the perturbated velocities and the growth rates vary with respect to different model parameters.

## 1. Introduction

In many biological processes, such as development, regeneration, and wound healing, epithelial cells move collectively[1] by cross-talk through cell-cell adhesions, cell-matrix interactions, and active intracellular contractilities[2]. Despite the debate on the chemical and molecular mechanisms involved in cell coordinated behaviors, it is recognized that mechanical factors are intimately related to these behaviors[3], as revealed by recently emerged technologies[4,5]. Among these mechanical factors, the material properties of epithelial tissues undoubtably play crucial roles in collective cell behaviors[6,7]. In epithelial morphogenesis, cells collectively move and sort as an active fluid or viscoelastic continuum[8]. Such movement can be globally oriented, driven by locally polarized individual cells[9]. Many theoretical models have been developed to study such movements[3]. These models can be categorized into two kinds: mesoscale models, including Vertex model[10], Voronoi model[11] and Cellular Potts model[12]; the other class is the continuum models, which encompass the phase-field model[13] and the active polar fluid/gel model[14].

During recent decades, many efforts have been devoted to the characterization of collective motions by using the active polar fluid/gel model. These works are not restricted to the study of epithelial tissues but also bacteria[15] and subcellular structures[16]. Here, due to the limitations in article length, we report a few representative works to grasp the great progress made in recent years in this field. In 2006, Kung and Marchetti developed the full theory for the hydrodynamics of polar liquid crystals[17], which is different from nematic liquid crystals. Voituriez et al.[18] applied the active polar gel model to the study of the actin filaments of eukaryotic cells. These theories, thereafter, have been applied to study the morphology of biofilms[19], the shape instability of crawling cells due to actin polymerization[20], the spatiotemporal order in geometrically confined bacterial active matter[21], and topological defects in tissue morphogenesis[22]. The active polar fluid/gel model has also been widely applied to the study of epithelial spreading. In 2017, Blanch-Mercader et al.[23] proposed a solvable active polar fluid model and applied it to the interoperations of force microscopy data from a variety of experiments. Later, he and Casademunt studied the hydrodynamic instabilities of the spreading epithelium[24]. They found that the combination of active contractility and traction force can lead to a nonlinear traveling wave in the epithelial sheet. Two years later, Moitrier et al.[25] adopted the model to estimate the relative material parameters for different cell lines. In the same year, Alert et al.[26] showed that the spreading epithelium undergoes active fingering instability by modeling the epithelium as a compressible viscous polar fluid. Perez-Gonzales et al.[27] adopted the same model to study the active wetting of epithelial tissues. The wetting transition comes from the competition between the intercellular contraction and the extracellular traction force. Recently, Trenado and her colleagues systematically analyzed the roles of cell polarization, substrate friction, and contractile stress played in the instabilities of the spreading epithelial front by the same model[28]. Their conclusion suggests that cell-substrate friction should be the main mechanism responsible for the fingering instability at the epithelial edge.

Among the many collective behaviors of epithelial cells, the wound healing/gap closure process has been well investigated from both the experimental and theoretical aspects[29]. It was found that the coordinated motions of cells lead to an improvement of the wound healing process[30]. Experimental works have revealed two major mechanisms for wound healing/gap closure[31]: the assembly of a supracellular actomyosin ring that generates purse-string contraction[5,32,33] and cell crawling through lamellipodial protrusions [34–36]. The morphogenic signals released by wounds also play a role in wound healing through chemotaxis[37]. Many theoretical models have been proposed to simulate the wound healing process at the tissue level. Here, we address a few advances made on the theoretical modeling on this topic during recent years, even though they are not necessarily related to our work. In 2014, Cochet-Escartin et al.[38] proposed a physical model that incorporates the border force, fiction, and tissue rheology. Wu and Amar studied the re-epithelialization driven by chemotaxis for a circular wound[39] and later studied wound healing by a model of growth mechanics[40]. In 2015, Valero et al.[41] worked out a wound healing model that is focused on material anisotropy and tissue proliferation. In 2016, Tepole developed a wound healing model that takes remodeling of tissue configuration, reaction-diffusion of contraction signals, and cellular mass conservation into account, bridging the junction between mechanics and system biology[42]. Recently, Roldan et al.[43] worked on the simulation of the formation of the supracellular actin ring by focusing on calcium ion transport in the cell sheet. Chojowski et al.[44] proposed an elastic phase-field approach to study the migration of the cell monolayer under the wound healing assay.

In this article, we study the planar epithelial monolayer with a circular migration front that encloses a gap in the center of the monolayer, as pictured in Fig. 1. In Section 2, we model the epithelial monolayer as a layer of compressible active polar fluid. We find that this model is analytically solvable under two special conditions by adopting an axisymmetric assumption. We then solve the model with the given boundary conditions. In Section 3, we study the instabilities of the propagating front of the spreading epithelium. In Section 4, we investigate the numerical properties of the solutions and discuss the velocity of the spreading front. We apply the solution to solve for a circular gap closure process and finally study the loss of the circularity of the morphology of the spreading front.

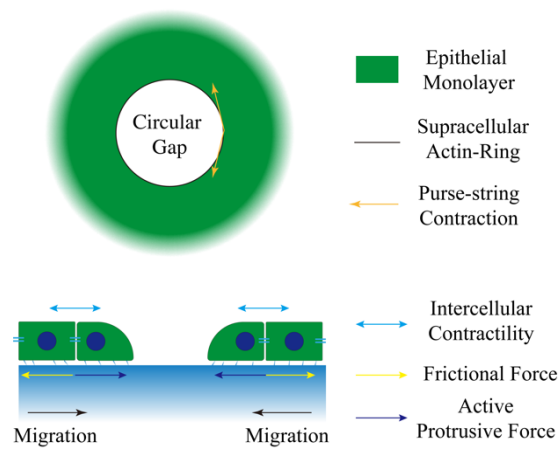

**Fig. 1.** Schematic sketch of the spreading epithelial monolayer with a circular migrating front that mimics the wound healing process. The top view: a supracellular actin ring generates a purse-string contraction; the side view: the epithelial monolayer

migrates towards the gap center under the combined effect of intercellular contractility, friction and active protrusive force. Forces are represented by arrows with different colors.

## 2. Theory

In particular, we consider a monolayer cell sheet adherent on a planar substrate with a circular gap located in its center. The cells propagate towards the gap center, aiming to close it by active protrusive forces and purse-string contraction. The protrusive forces are generated by the cell actomyosin cytoskeleton and transmitted to the substrate through focal adhesions, while the purse-string contraction is developed in a supracellular cellular actomyosin ring. The combination of the two forces enables cells to migrate collectively, as shown in Fig. 1. To characterize this collective motion, we model the cell monolayer as a compressible active polar fluid, which is described in terms of the two-dimensional depth-averaged polar field $\mathbf{p}(\tilde{\mathbf{r}}, \tilde{t})$ and the velocity field $\tilde{\mathbf{v}}(\tilde{\mathbf{r}}, \tilde{t})$, with $\tilde{\mathbf{r}}$ and $\tilde{t}$ being the position vector and time, respectively. We assume that the free energy of the active polar fluid consists of the Landau-de Gennes part around the isotropic state $\mathbf{p} = \mathbf{0}$ and a gradient term raised from polar elasticity[26,45]:

$$F = \iint [\frac{G}{2}|\mathbf{p}|^2 + \frac{\lambda}{2}\nabla\mathbf{p}:\nabla\mathbf{p}]\, d^2\tilde{r}, \tag{1}$$

where $G$ is a restoring coefficient and $\lambda$ is the one-constant approximation of the Frank elasticity constant. By neglecting the polarity advection, corotation and the flow alignment effect, we write the polarity dynamics as:

$$\frac{\partial \mathbf{p}}{\partial \tilde{t}} = -L\frac{\partial F}{\partial \mathbf{p}}, \tag{2}$$

where $L$ is the mobility parameter. In the case that the polarity dynamics is much faster than the cell spreading dynamics, $\partial_{\tilde{t}}\mathbf{p} \approx \mathbf{0}$, and thus, Eq. (2) is always at equilibrium. Under this assumption, the polarity field is given by:

$$L_c^2 \nabla^2 \mathbf{p} = \mathbf{p}, \tag{3}$$

with the polar length scale $L_c = \sqrt{\lambda/G}$ that depicts the decay of the cellular polarity field. The total stress $\tilde{\boldsymbol{\sigma}}$ within the tissue is composed of three parts:

$$\tilde{\boldsymbol{\sigma}} = \tilde{\boldsymbol{\sigma}}^s + \tilde{\boldsymbol{\sigma}}^a + \tilde{\boldsymbol{\sigma}}^e. \tag{4}$$

The first term in the total stress is the symmetric part of the deviatoric stress and reads:

$$\tilde{\boldsymbol{\sigma}}^s = \eta\, dev(\nabla\tilde{\mathbf{v}}) + \bar{\eta}\, tr(\nabla\tilde{\mathbf{v}})\mathbf{I} + \frac{\nu}{2}[\mathbf{p}\otimes\mathbf{h} + \mathbf{h}\otimes\mathbf{p} - tr(\mathbf{p}\otimes\mathbf{h})\mathbf{I}] + \nu' tr(\mathbf{p}\otimes\mathbf{h})\mathbf{I} -$$

$$\zeta_0\left[\mathbf{p}\otimes\mathbf{p} - \frac{1}{2}tr(\mathbf{p}\otimes\mathbf{p})\mathbf{I}\right] - \zeta_1 tr(\mathbf{p}\otimes\mathbf{p})\mathbf{I} - \zeta_2 \mathbf{I}, \tag{5}$$

where $\mathbf{h}$ equals $-\frac{\partial F}{\partial \mathbf{p}}$. The second term $\tilde{\boldsymbol{\sigma}}^a$ corresponds to the asymmetric part and is written as:

$$\tilde{\boldsymbol{\sigma}}^a = \frac{1}{2}(\mathbf{p}\otimes\mathbf{h} - \mathbf{h}\otimes\mathbf{p}). \tag{6}$$

The third term $\tilde{\boldsymbol{\sigma}}^e$ stands for the Ericksen stress:

$$\tilde{\boldsymbol{\sigma}}^e = -p\mathbf{I} - \lambda(\nabla\mathbf{p})^2, \tag{7}$$

where $p$ is the pressure caused by the change in the cell number density. By letting $\bar{\eta} = \eta/2$, $v' = v/2$, $\zeta_1 = \zeta_0/2$, and $\zeta_2 = 0$ and noticing $\mathbf{h} = 0$, we have:

$$\tilde{\boldsymbol{\sigma}} = sym(\nabla\tilde{\mathbf{v}})\eta - \zeta_0 \mathbf{p}\otimes\mathbf{p} - p\mathbf{I} - \lambda(\nabla\mathbf{p})^2. \tag{8}$$

By neglecting the cell proliferation and the second-order terms on $\nabla\mathbf{p}$, the total stress can be further simplified as:

$$\tilde{\boldsymbol{\sigma}} = sym(\nabla\tilde{\mathbf{v}})\eta - \zeta_0 \mathbf{p}\otimes\mathbf{p}. \tag{9}$$

The mechanical equilibrium is described through the balance between the cell monolayer stress $\tilde{\boldsymbol{\sigma}}$ and the extracellular traction force $\mathbf{T}$:

$$h_c \nabla \cdot \tilde{\boldsymbol{\sigma}} + \mathbf{T} = \mathbf{0}, \tag{10}$$

where $h_c$ is the monolayer thickness. The extracellular traction force $\mathbf{T}$ consists of two parts: the cell-substrate friction force and the active protrusive force related to the cell polarity:

$$\mathbf{T} = -\xi\tilde{\mathbf{v}} + T_0 \mathbf{p}. \tag{11}$$

The parameters $\xi$ and $T_0$ are the friction coefficient and the active protrusive force coefficient.

We next introduce the following way to nondimensionalize the system.

$$\mathbf{r} = \frac{\tilde{\mathbf{r}}}{L_c}, \quad \mathbf{v} = \frac{\tilde{\mathbf{v}}\eta}{L_c T_0}, \quad \boldsymbol{\sigma} = \frac{\tilde{\boldsymbol{\sigma}}}{T_0}. \tag{12}$$

The equilibrium of the polarity field $\mathbf{p}(\mathbf{r},t)$ now reads:

$$\nabla^2 \mathbf{p} = \mathbf{p}. \tag{13}$$

In the axisymmetric case, it can be written in the component form:

$$\left[\frac{\partial^2}{\partial r^2} + \frac{1}{r}\frac{\partial}{\partial r} - \left(1 + \frac{1}{r^2}\right)\right] p_r = 0, \tag{14a}$$

$$\left[\frac{\partial^2}{\partial r^2} + \frac{1}{r}\frac{\partial}{\partial r} - \left(1 + \frac{1}{r^2}\right)\right] p_\theta = 0, \tag{14b}$$

where $r$ is the dimensionless radial coordinate and $p_r$ and $p_\theta$ are the radial and tangential components of $\mathbf{p}(\mathbf{r},\tilde{t})$, respectively. The solution to the above equations is:

$$p_r = aI_1(r) + bK_1(r), \tag{15a}$$
$$p_\theta = cI_1(r) + dK_1(r), \tag{15b}$$

where $I_1$ and $K_1$ are the modified Bessel functions of the first and second kinds; $a$, $b$, $c$, and $d$ are undetermined constants. The dimensionless mechanical equilibrium in the radial direction is:

$$\frac{\partial \sigma_{rr}}{\partial r} + \frac{\sigma_{rr} - \sigma_{\theta\theta}}{r} + f_r = 0. \tag{16}$$

The dimensionless constitutive equation is:

$$\boldsymbol{\sigma} = \frac{1}{2}[\nabla\mathbf{v} + (\nabla\mathbf{v})^T] - \zeta \mathbf{p}\otimes\mathbf{p}, \tag{17}$$

with $\zeta = \zeta_0/T_0$. The dimensionless body force is written as:

$$f_r = -\alpha^2 v_r + \beta p_r, \tag{18}$$

where $\alpha^2 = \frac{\xi L_c^2}{\eta h_c}$, $\beta = \frac{L_c}{h_c}$. By substituting the constitutive law into the mechanical equilibrium in the radial direction, we obtain the mechanical equilibrium in the velocity

form:
$$\left[\frac{\partial^2}{\partial r^2}+\frac{1}{r}\frac{\partial}{\partial r}-\left(\alpha^2+\frac{1}{r^2}\right)\right]v_r = -\beta p_r + \zeta\left(\frac{p_r^2-p_\theta^2}{r}+2p_r\frac{\partial p_r}{\partial r}\right). \quad (19)$$

This equation can be solved analytically when $\alpha \neq 1$. The specific solution to the equation can be assumed to have the following form:
$$v_r^1 = AI_0(r)I_1(r)+BK_0(r)K_1(r)+CI_0(r)K_1(r)+DK_0(r)I_1(r)+EI_1(r)+FK_1(r), \quad (20)$$
where $A\sim F$ are constants that need to be determined. The general solution, in contrast, is:
$$v_r^0 = GI_1(\alpha r) + HK_1(\alpha r). \quad (21)$$

We substitute the specific solution to the velocity field and the solutions to the polarity field back into the mechanical equilibrium and compare the terms on both sides of the equation:
$$\begin{cases}(2-\alpha^2)A = (a^2-c^2)\zeta, -(2-\alpha^2)B = (b^2-d^2)\zeta \\ (4-\alpha^2)A = 2a^2\zeta, -(4-\alpha^2)B = 2b^2\zeta \\ (2-\alpha^2)(C-D) = 2(ab-cd)\zeta \\ -(2-\alpha^2)D + 2C = 2ab\zeta, (2-\alpha^2)C - 2D = 2ab\zeta \\ (1-\alpha^2)E = -a\beta, (1-\alpha^2)F = -b\beta\end{cases}. \quad (22)$$

It can be deduced that:
$$\begin{cases}ac(bc-ad) = 0, bd(bc-ad) = 0 \\ a^2d^2 = b^2c^2, \alpha^2ab = (4-\alpha^2)cd \\ C = -D\end{cases}. \quad (23)$$

The mechanical equilibrium in the tangential direction is:
$$\left[\frac{\partial^2}{\partial r^2}+\frac{1}{r}\frac{\partial}{\partial r}-\left(2\alpha^2+\frac{1}{r^2}\right)\right]v_\theta = -2\beta p_\theta + 2\zeta\left(\frac{2p_r p_\theta}{r}+\frac{\partial p_r}{\partial r}p_\theta + p_r\frac{\partial p_\theta}{\partial r}\right). \quad (24)$$

This equation can be solved analytically when $\alpha \neq 1/\sqrt{2}$ by assuming the general solution with the same form from Eq. (20):
$$v_\theta^1 = \tilde{A}I_0(r)I_1(r)+\tilde{B}K_0(r)K_1(r)+\tilde{C}I_0(r)K_1(r)+\tilde{D}K_0(r)I_1(r)+\tilde{E}I_1(r)+\tilde{F}K_1(r), \quad (25)$$
where $\tilde{A}\sim\tilde{F}$ are constants that need to be determined. The general solution, in contrast, is:
$$v_\theta^0 = \tilde{G}I_1(\sqrt{2}\alpha r) + \tilde{H}K_1(\sqrt{2}\alpha r). \quad (26)$$

Similarly, we substitute the specific solution and the solutions to the polarity field back into the mechanical equilibrium and compare the terms on both sides of the equation:
$$\begin{cases}(1-\alpha^2)\tilde{A} = 2ac\zeta, -(1-\alpha^2)\tilde{B} = 2bd\zeta \\ (2-\alpha^2)\tilde{A} = 2ac\zeta, -(2-\alpha^2)\tilde{B} = 2bd\zeta \\ (1-\alpha^2)\tilde{C} = (ad+bc)\zeta, (2-\alpha^2)\tilde{C} = (ad+bc)\zeta, \tilde{C} = -\tilde{D} \\ (1-2\alpha^2)\tilde{E} = -2c\beta, (1-2\alpha^2)\tilde{F} = -2d\beta\end{cases}. \quad (27)$$

It can be deduced that:
$$\begin{cases}ac = bd = 0 \\ ad + bc = 0 \\ \tilde{A} = \tilde{B} = \tilde{C} = \tilde{D} = 0\end{cases}. \quad (28)$$

By considering Eq. (22), Eq. (23), Eq. (27), and Eq. (28), we can find the two possible nonzero solutions for $\{p_r, p_\theta\}$. The first solution is derived when $c = 0$, $d = 0$. This further leads to $\alpha = 0$, which means that the cell-substrate friction is negligible compared with the viscosity of the tissue. Now back substituting this into Eq. (22), we obtain the first set of analytical solutions:

$$\begin{cases} 2A = a^2\zeta, 2B = -b^2\zeta, C = ab\zeta \\ E = -a\beta, F = -b\beta \\ \tilde{A} = \tilde{B} = \tilde{C} = \tilde{E} = \tilde{F} = 0 \end{cases}. \tag{29}$$

The solution for the polarity is:
$$p_r = aI_1(r) + bK_1(r), \tag{30a}$$
$$p_\theta = 0. \tag{30b}$$

The solution for the velocity is:
$$v_r = \frac{a^2\zeta I_0(r)I_1(r) - b^2\zeta K_0(r)K_1(r) + ab\zeta[I_0(r)K_1(r) - K_0(r)I_1(r)]}{2}$$
$$-a\beta I_1(r) - b\beta K_1(r) + Gr + H/r, \tag{31a}$$
$$v_\theta = \tilde{G}r + \tilde{H}/r. \tag{31b}$$

The second possible set of analytical solutions is obtained when $a = 0$, $b = 0$, which further leads to $\alpha = 2$. Similar to Eq. (29), one can derive the following:

$$\begin{cases} 2A = c^2\zeta, 2B = -d^2\zeta, 2C = cd\zeta \\ \tilde{A} = \tilde{B} = \tilde{C} = E = F = 0 \\ \tilde{E} = \frac{2c\beta}{7}, \tilde{F} = \frac{2d\beta}{7} \end{cases}. \tag{32}$$

The solution for the polarity becomes:
$$p_r = 0, \tag{33a}$$
$$p_\theta = cI_1(r) + dK_1(r). \tag{33b}$$

The solution for the velocity hence reads:
$$v_r = \frac{c^2\zeta I_0(r)I_1(r) - d^2\zeta K_0(r)K_1(r) + cd\zeta[I_0(r)K_1(r) - K_0(r)I_1(r)]}{2} + GI_1(2r) + HK_1(2r), \tag{34a}$$

$$v_\theta = \frac{2c\beta}{7}I_1(r) + \frac{2d\beta}{7}K_1(r) + \tilde{G}I_1(2\sqrt{2}r) + \tilde{H}K_1(2\sqrt{2}r). \tag{34b}$$

The constant can be determined by specifying particular boundary conditions onto the polarity field, velocity field or stress field. Here. Let us consider a special case in which the size of the cell monolayer is much larger than the radius of the circular gap $R_{in}$. We assume that both $p_{r,\theta}$ and $v_{r,\theta}$ vanish at the place that is far away from the gap. Such boundary conditions can be written as $p_{r,\theta}(+\infty), v_{r,\theta}(+\infty) = 0$. One could immediately have that $a = c = 0$. If $\alpha \to 0$, we can immediately have:

$$p_r = bK_1(r), \tag{35a}$$
$$p_\theta = 0, \tag{35b}$$
$$v_r = -\frac{b^2\zeta K_0(r)K_1(r)}{2} - b\beta K_1(r) + H/r, \tag{35c}$$
$$v_\theta = \tilde{H}/r. \tag{35d}$$

At $r = R_{in}$, the purse-string contraction induces a boundary traction that equals $\gamma\kappa\mathbf{n}$, where $\gamma$ is the depth-averaged dimensionless purse-string contraction generated by the supracellular contractile ring at the gap boundary. The term $\gamma$ is assumed to be dependent on the dimensionless ring curvature $\kappa$ through the power law: $\gamma = \gamma_m |\kappa|^m$,[33,] where the surface tension coefficient $\gamma_n$ is nondimensionalized by $\gamma_m = \Gamma_m/T_0(L_c)^{m+1}$, with $\Gamma_m$ being the surface tension coefficient with dimensions. Since we neglect cell proliferation, we could assume that the supracellular contractile ring is always at the cell front and overlaps with the gap boundary. The curvature of the contractile ring is described by $\kappa$, and **n** stands for the outer unit normal of the gap boundary, which points toward the gap center. For a circular gap, $\kappa = -1/R_{in}$; hence, the boundary condition is $\sigma_{rr}|_{r=R_{in}} = \frac{\gamma(R_{in})}{R_{in}}$, $\sigma_{r\theta}|_{r=R_{in}} = 0$, and $|\mathbf{p}| = 1$.

Then, the constants are calculated as:

$$b^2 = \frac{1}{K_1^2(R_{in})}, \tag{36a}$$

$$H = -R_{in}^2 \left\{ \frac{\zeta[K_0'(R_{in})K_1(R_{in})+K_0(R_{in})K_1'(R_{in})]}{2K_1^2(R_{in})} + b\beta K_1'(R_{in}) + \zeta \right\} - \gamma(R_{in})R_{in}, \tag{36b}$$

$$\tilde{H} = 0. \tag{36c}$$

Let the polarity field point towards the center of the gap be $p_r = -K_1(r)/K_1(R_{in})$; the velocity at $r = R_{in}$ equals:

$$v_r(R_{in}) = -\beta R_{in} \frac{K_0(R_{in})}{K_1(R_{in})} + R_{in}\zeta \frac{K_0^2(R_{in})}{2K_1^2(R_{in})} - \frac{R_{in}\zeta}{2} - \gamma(R_{in}). \tag{37}$$

If $\alpha = 2$, similarly, one could derive:

$$p_r = 0, \tag{38a}$$
$$p_\theta = dK_1(r), \tag{38b}$$
$$v_r = -\frac{d^2\zeta K_0(r)K_1(r)}{2} + HK_1(2r), \tag{38c}$$
$$v_\theta = \frac{2d\beta}{7}K_1(r) + \tilde{H}K_1(2\sqrt{2}r). \tag{38d}$$

At $r = R_{in}$, we impose the same boundary conditions that read $\sigma_{rr}|_{r=R_{in}} = \frac{\gamma(R_{in})}{R_{in}}$, $\sigma_{r\theta}|_{r=R_{in}} = 0$, and $|\mathbf{p}| = 1$ at the gap boundary. If we impose $d > 0$, the constants are then derived as:

$$d^2 = \frac{1}{K_1^2(R_{in})}, \tag{39a}$$

$$H = \frac{\zeta[K_0'(R_{in})K_1(R_{in})+K_0(R_{in})K_1'(R_{in})]}{4K_1'(2R_{in})K_1^2(R_{in})} + \frac{\gamma(R_{in})}{2R_{in}K_1'(2R_{in})}, \tag{39b}$$

$$\tilde{H} = \frac{2\beta[R_{in}K_1'(R_{in})-K_1(R_{in})]}{7K_1(R_{in})[K_1(2\sqrt{2}R_{in})-2\sqrt{2}R_{in}K_1'(2\sqrt{2}R_{in})]}. \tag{39c}$$

The radial velocity at $r = R_{in}$ equals:

$$v_r = \frac{\zeta\{2K_0(R_{in})K_1(R_{in})K_0(2R_{in})-K_1(2R_{in})[K_0^2(R_{in})+K_1^2(R_{in})]\}}{4K_1'(2R_{in})K_1^2(R_{in})} + \frac{\gamma(R_{in})K_1(2R_{in})}{2R_{in}K_1'(2R_{in})}. \tag{40}$$

The two sets of analytical solutions suggest two possible spreading modes for a circular front. The first mode is that all the cells are polarized in the radial direction and migrate towards the gap centre with the help of purse-string contraction and active traction force. The second mode is that all the cells are polarized in the tangential direction and move in both directions. This kind of collective motion is promoted by the active contractility in the tangential direction and the purse-string contraction.

### 3. Linear stability analysis

In this section, we investigate the loss of morphological stability of the epithelial front during collective migration. First, we allow the loss of axisymmetry in both the equilibrium in the polarity field and the stress field:

$$\left[\frac{\partial^2}{\partial r^2} + \frac{1}{r}\frac{\partial}{\partial r} - \left(1 + \frac{1}{r^2}\right)\right] p_r + \frac{1}{r^2}\partial_\theta^2 p_r - \frac{2}{r^2}\partial_\theta p_\theta = 0, \tag{41a}$$

$$\left[\frac{\partial^2}{\partial r^2} + \frac{1}{r}\frac{\partial}{\partial r} - \left(1 + \frac{1}{r^2}\right)\right] p_\theta + \frac{1}{r^2}\partial_\theta^2 p_\theta + \frac{2}{r^2}\partial_\theta p_r = 0, \tag{41b}$$

$$\frac{\partial \sigma_{rr}}{\partial r} + \frac{1}{r}\frac{\partial}{\partial \theta}\sigma_{\theta r} + \frac{\sigma_{rr} - \sigma_{\theta\theta}}{r} + f_r = 0, \tag{41c}$$

$$\frac{\partial \sigma_{r\theta}}{\partial r} + \frac{1}{r}\frac{\partial}{\partial \theta}\sigma_{\theta\theta} + \frac{\sigma_{r\theta} + \sigma_{\theta r}}{r} + f_\theta = 0, \tag{41d}$$

Next, we introduce a small perturbation to the morphology of the gap, the polarity field, and the velocity field:

$$R_{in}(\theta, t) = R_{in}^0 + \delta R(\theta, t), \tag{42a}$$
$$p_r(r, \theta, t) = p_r^0(r) + \delta p_r(r, \theta, t), \tag{42b}$$
$$p_\theta(r, \theta, t) = p_\theta^0(r) + \delta p_\theta(r, \theta, t), \tag{42c}$$
$$v_r(r, \theta, t) = v_r^0(r) + \delta v_r(r, \theta, t), \tag{42d}$$
$$v_\theta(r, \theta, t) = v_\theta^0(r) + \delta v_\theta(r, \theta, t). \tag{42e}$$

Based on the constitutive equation Eq. (17), the perturbations of the stress components then become:

$$\delta \sigma_{rr} = \frac{\partial \delta v_r}{\partial r} - 2\zeta p_r^0 \delta p_r, \tag{43a}$$

$$\delta \sigma_{\theta\theta} = \frac{\delta v_r}{r} + \frac{\partial \delta v_\theta}{r\partial \theta} - 2\zeta p_\theta^0 \delta p_\theta, \tag{43b}$$

$$\delta \sigma_{r\theta} = \delta \sigma_{\theta r} = \frac{1}{2}\left(\frac{\partial \delta v_\theta}{\partial r} - \frac{\delta v_\theta}{r} + \frac{\partial \delta v_r}{r\partial \theta}\right) - \zeta(p_r^0 \delta p_\theta + p_\theta^0 \delta p_r). \tag{43c}$$

The perturbated equations of both the polarity and stress field become:

$$\left[\frac{\partial^2}{\partial r^2} + \frac{1}{r}\frac{\partial}{\partial r} - \left(1 + \frac{1}{r^2}\right)\right] \delta p_r + \frac{1}{r^2}\partial_\theta^2 \delta p_r - \frac{2}{r^2}\partial_\theta \delta p_\theta = 0, \tag{44a}$$

$$\left[\frac{\partial^2}{\partial r^2} + \frac{1}{r}\frac{\partial}{\partial r} - \left(1 + \frac{1}{r^2}\right)\right] \delta p_\theta + \frac{1}{r^2}\partial_\theta^2 \delta p_\theta + \frac{2}{r^2}\partial_\theta \delta p_r = 0, \tag{44b}$$

$$\frac{\partial \delta \sigma_{rr}}{\partial r} + \frac{1}{r}\frac{\partial \delta \sigma_{\theta r}}{\partial \theta} + \frac{\delta \sigma_{rr} - \delta \sigma_{\theta\theta}}{r} - \alpha^2 \delta v_r + \beta \delta p_r = 0, \tag{44c}$$

$$\frac{\partial \delta \sigma_{r\theta}}{\partial r} + \frac{1}{r}\frac{\partial \delta \sigma_{\theta\theta}}{\partial \theta} + \frac{\delta \sigma_{r\theta} + \delta \sigma_{\theta r}}{r} - \alpha^2 \delta v_\theta + \beta \delta p_\theta = 0. \tag{44d}$$

To solve the above perturbation equations, we need to find the boundary conditions in the perturbated configuration. We specify the boundary condition at the perturbated gap boundary $R = R_{in}^0 + \delta R$ to be:

$$p_r^2(R, \theta, t) + p_\theta^2(R, \theta, t) = 1, \tag{45a}$$

$$\boldsymbol{\sigma}(R,\theta,t)\mathbf{n}(R,\theta,t) = -\gamma\kappa(R,\theta,t)\mathbf{n}(R,\theta,t). \tag{45b}$$

The outer unit normal in the perturbated configuration becomes:

$$\mathbf{n}(R,\theta,t) \approx -\mathbf{e}_r + \frac{1}{R_{in}^0}\frac{d\delta R}{d\theta}\mathbf{e}_\theta. \tag{46}$$

Let us perturbate the system from the state that $\alpha = 0$, $p_\theta^0 = 0$, and $v_\theta^0 = 0$. The perturbated polarity field has the following connections with the unperturbed one:

$$p_r(R,\theta,t) = p_r^0(R_{in}^0) + \partial_r p_r^0(R_{in}^0)\delta R + \delta p_r(R,\theta,t), \tag{47a}$$
$$p_\theta(R,\theta,t) = \delta p_\theta(R,\theta,t). \tag{47b}$$

By plugging the above relations into Eq. (45a) and neglecting the second-order terms in perturbations, one can derive the boundary condition for $\delta p_r$:

$$\delta p_r + \frac{\partial p_r^0(R_{in}^0)}{\partial r}\delta R = 0. \tag{48}$$

The curvature can be calculated through the following equation in the polar coordinate system:

$$\kappa(R,\theta,t) = -\frac{1 + 2(\partial_\theta \delta R/R)^2 - \partial_\theta^2 \delta R/R}{R[1+(\partial_\theta \delta R/R)^2]^{3/2}}, \tag{49}$$

which can be approximated as:

$$\kappa(R,\theta,t) \approx -\frac{1}{R}\left[1 + 2\left(\frac{\partial_\theta \delta R}{R}\right)^2 - \frac{\partial_\theta^2 \delta R}{R}\right]\left[1 - \frac{3}{2}\left(\frac{\partial_\theta \delta R}{R}\right)^2\right]$$

$$\approx -\frac{1}{R}\left(1 - \frac{1}{R}\frac{\partial^2 \delta R}{\partial \theta^2}\right) = -\frac{1}{R_{in}^0\left(1+\frac{\delta R}{R_{in}^0}\right)}\left[1 - \frac{1}{R_{in}^0\left(1+\frac{\delta R}{R_{in}^0}\right)}\frac{\partial^2 \delta R}{\partial \theta^2}\right]$$

$$\approx -\frac{1}{R_{in}^0}\left(1 - \frac{\delta R}{R_{in}^0}\right)\left[1 - \frac{1}{R_{in}^0}\frac{\partial^2 \delta R}{\partial \theta^2}\left(1 - \frac{\delta R}{R_{in}^0}\right)\right]$$

$$\approx -\frac{1}{R_{in}^0}\left(1 - \frac{\delta R}{R_{in}^0} - \frac{1}{R_{in}^0}\frac{\partial^2 \delta R}{\partial \theta^2}\right). \tag{50}$$

The perturbated stress field, similar to the polarity field, has the following connections with the unperturbed one:

$$\sigma_{rr}(R,\theta,t) = \sigma_{rr}^0(R_{in}^0) + \partial_r\sigma_{rr}^0(R_{in}^0)\delta R + \delta\sigma_{rr}(R,\theta,t), \tag{51a}$$
$$\sigma_{\theta\theta}(R,\theta,t) = \sigma_{\theta\theta}^0(R_{in}^0) + \partial_r\sigma_{\theta\theta}^0(R_{in}^0)\delta R + \delta\sigma_{\theta\theta}(R,\theta,t), \tag{51b}$$
$$\sigma_{r\theta}(R,\theta,t) = \sigma_{\theta r}(R,\theta,t) = \delta\sigma_{r\theta}(R,\theta,t). \tag{51c}$$

By substituting Eq. (51a) ~ (51c), Eq. (50), and Eq. (46) into Eq. (45b) and omitting high-order terms in perturbations, we derive:

$$\delta\sigma_{rr} = -\left[\frac{\partial\sigma_{rr}^0(R_{in}^0)}{\partial r} + \frac{(m+1)\gamma_m}{(R_{in}^0)^{m+2}}\right]\delta R - \frac{(m+1)\gamma_m}{(R_{in}^0)^{m+2}}\frac{d^2\delta R}{d\theta^2}, \tag{52a}$$

$$\delta\sigma_{r\theta} = \left[\sigma_{\theta\theta}^0(R_{in}^0) - \frac{\gamma_m}{(R_{in}^0)^{m+1}}\right]\frac{1}{R_{in}^0}\frac{d\delta R}{d\theta}. \tag{52b}$$

The perturbations can be expanded by trigonometric series:

$$\delta R(\theta,t) = \sum_{n=0}^{\infty} \delta\tilde{R}_n(t)\cos(n\theta), \tag{53a}$$
$$\delta p_r(r,\theta,t) = \sum_{n=0}^{\infty} \delta\tilde{p}_{r,n}(r,t)\cos(n\theta), \tag{53b}$$
$$\delta p_\theta(r,\theta,t) = \sum_{n=0}^{\infty} \delta\tilde{p}_{\theta,n}(r,t)\sin(n\theta), \tag{53c}$$

$$\delta v_r(r,\theta,t) = \sum_{n=0}^{\infty} \delta \tilde{v}_{r,n}(r,t)\cos(n\theta), \tag{53d}$$
$$\delta v_\theta(r,\theta,t) = \sum_{n=0}^{\infty} \delta \tilde{v}_{\theta,n}(r,t)\sin(n\theta). \tag{53e}$$

In terms of the Fourier modes, the equations of the polarity field are as follows:

$$\left[\frac{\partial^2}{\partial r^2} + \frac{1}{r}\frac{\partial}{\partial r} - \left(1 + \frac{1+n^2}{r^2}\right)\right]\delta\tilde{p}_{r,n} = \frac{2n}{r^2}\delta\tilde{p}_{\theta,n}, \tag{54a}$$

$$\left[\frac{\partial^2}{\partial r^2} + \frac{1}{r}\frac{\partial}{\partial r} - \left(1 + \frac{1+n^2}{r^2}\right)\right]\delta\tilde{p}_{\theta,n} = \frac{2n}{r^2}\delta\tilde{p}_{r,n}. \tag{54b}$$

The mechanical equilibrium in the radial and tangential directions then reads:

$$\left(\frac{\partial^2}{\partial r^2} + \frac{1}{r}\frac{\partial}{\partial r} - \frac{1+\frac{n^2}{2}}{r^2}\right)\delta\tilde{v}_{r,n} + \frac{n}{2r}\left(\frac{\partial}{\partial r} - \frac{3}{r}\right)\delta\tilde{v}_{\theta,n}$$
$$+[\beta - \zeta(2p_r^0\frac{\partial}{\partial r} + 2\frac{\partial p_r^0}{\partial r} - 2\frac{p_r^0}{r})]\delta\tilde{p}_{r,n} - \frac{n\zeta}{r}p_r^0\delta\tilde{p}_{\theta,n} = 0, \tag{55a}$$

$$\frac{n}{2r}\left(-3\frac{\partial}{\partial r} + \frac{1}{r}\right)\delta\tilde{v}_{r,n} + \frac{1}{2}\left(\frac{\partial^2}{\partial r^2} + \frac{1}{r}\frac{\partial}{\partial r} - \frac{1+2n^2}{r^2} - 2\alpha^2\right)\delta\tilde{v}_{\theta,n}$$
$$+[\beta - \zeta(\frac{\partial p_r^0}{\partial r} + 2\frac{p_r^0}{r} + p_r^0\frac{\partial}{\partial r})]\delta\tilde{p}_{\theta,n} = 0. \tag{55b}$$

The boundary conditions in the trigonometric modes become:

$$\delta\tilde{p}_{r,n} + \frac{\partial p_r^0(R_{in}^0)}{\partial r}\delta\tilde{R}_n = 0, \tag{56a}$$

$$\delta\tilde{\sigma}_{rr,n} = -[\frac{\partial \sigma_{rr}^0(R_{in}^0)}{\partial r} + \frac{\gamma_m(1-n^2)(m+1)}{(R_{in}^0)^{m+2}}]\delta\tilde{R}_n, \tag{56b}$$

$$\delta\tilde{\sigma}_{r\theta,n} = -n[\sigma_{\theta\theta}^0(R_{in}^0) - \frac{\gamma_m}{(R_{in}^0)^{m+1}}]\frac{1}{R_{in}^0}\delta\tilde{R}_n. \tag{56c}$$

Similarly, we could also perturbate the system from the state that $\alpha = 2$, $p_r^0 = 0$. The perturbated polarity field has the following connections with the unperturbed one:
$$p_r(R,\theta,t) = \delta p_r(R,\theta,t), \tag{57a}$$
$$p_\theta(R,\theta,t) = p_\theta^0(R_{in}^0) + \partial_r p_\theta^0(R_{in}^0)\delta R + \delta p_\theta(R,\theta,t). \tag{57b}$$

By plugging the above relations into the former equation and neglecting the second-order terms in perturbations, one can derive the boundary condition for $\delta p_r$:

$$\delta p_\theta + \frac{\partial p_\theta^0(R_{in}^0)}{\partial r}\delta R = 0. \tag{58}$$

The perturbated stress field now has the following connections with the unperturbed one:
$$\sigma_{rr}(R,\theta,t) = \sigma_{rr}^0(R_{in}^0) + \partial_r\sigma_{rr}^0(R_{in}^0)\delta R + \delta\sigma_{rr}(R,\theta,t), \tag{59a}$$
$$\sigma_{\theta\theta}(R,\theta,t) = \sigma_{\theta\theta}^0(R_{in}^0) + \partial_r\sigma_{\theta\theta}^0(R_{in}^0)\delta R + \delta\sigma_{\theta\theta}(R,\theta,t), \tag{59b}$$
$$\sigma_{r\theta}(R,\theta,t) = \sigma_{\theta r}(R,\theta,t) = \sigma_{r\theta}^0(R_{in}^0) + \partial_r\sigma_{r\theta}^0(R_{in}^0)\delta R + \delta\sigma_{r\theta}(R,\theta,t), \tag{59c}$$

Similar to how one can obtain Eq. (52a) ~ (52b), the perturbated stress boundary condition for the case that $\alpha = 2$ reads:

$$\delta\sigma_{rr} = -\left[\frac{\partial \sigma_{rr}^0(R_{in}^0)}{\partial r} + \frac{(m+1)\gamma_m}{(R_{in}^0)^{m+2}}\right]\delta R - \frac{(m+1)\gamma_m}{(R_{in}^0)^{m+2}}\frac{d^2\delta R}{d\theta^2}, \tag{60a}$$

$$\delta\sigma_{r\theta} = \left[\sigma_{\theta\theta}^0(R_{in}^0) - \frac{\gamma_m}{(R_{in}^0)^{m+1}}\right]\frac{1}{R_{in}^0}\frac{d\delta R}{d\theta} - \frac{\partial\sigma_{r\theta}^0(R_{in}^0)}{\partial r}\delta R. \tag{60b}$$

From Eq. (60), we can see that the perturbation of velocity and the polarity field should be expanded as a combination of sine and cosine functions if $\delta R$ is expanded in the way of Eq. (53a):

$$\delta R(\theta, t) = \sum_{n=0}^{\infty} \delta\tilde{R}_n(t)\cos(n\theta), \tag{61a}$$

$$\delta p_r(r, \theta, t) = \sum_{n=0}^{\infty} \delta\tilde{p}_{r,n}^1(r, t)\cos(n\theta) + \delta\tilde{p}_{r,n}^2(r, t)\sin(n\theta), \tag{61b}$$

$$\delta p_\theta(r, \theta, t) = \sum_{n=0}^{\infty} \delta\tilde{p}_{\theta,n}^1(r, t)\cos(n\theta) + \delta\tilde{p}_{\theta,n}^2(r, t)\sin(n\theta), \tag{61c}$$

$$\delta v_r(r, \theta, t) = \sum_{n=0}^{\infty} \delta\tilde{v}_{r,n}^1(r, t)\cos(n\theta) + \delta\tilde{v}_{r,n}^2(r, t)\sin(n\theta), \tag{61d}$$

$$\delta v_\theta(r, \theta, t) = \sum_{n=0}^{\infty} \delta\tilde{v}_{\theta,n}^1(r, t)\cos(n\theta) + \delta\tilde{v}_{\theta,n}^2(r, t)\sin(n\theta), \tag{61e}$$

Before we approach the next section, we would like to discuss the biophysical background in which the two sets of analytical solutions can be achieved. The first set of analytical solutions requires $\alpha$ to be approximately zero. For an epithelial monolayer with cell thickness $h_c \approx 5$ μm and shear viscosity $\eta \sim 10^7$ Pa·s, if the nematic length $L_c$ and the cell-substrate coefficient $\xi$ are estimated as $L_c \approx 25$ μm and $\xi/h_c \approx 10$ Pa·s/μm$^2$, the value of $\alpha = (\frac{\xi L_c^2}{\eta h_c})^{1/2}$ can be calculated as $\alpha \approx 0.025 \ll 1$. This means that when the viscosity dominates over friction, the cell-substrate friction can be neglected, and $\alpha$ can be roughly considered zero. In contrast, if the cell viscosity is as small as $\eta \sim 10^5$ Pa·s, while the nematic length $L_c$ is estimated as $L_c \approx 10^2$ μm, then $\alpha$ can be estimated as $\alpha \approx 3.16$. This implies that when the effect of friction is comparable with that of viscosity, the condition that $\alpha = 2$ can be achieved.

**Table. 1.** The values of the physical and dimensionless parameters used in the article.

| Parameters | Definition | Values | Ref. |
|---|---|---|---|
| Physical | | | |
| $h_c$ | Monolayer thickness | 5 μm | Ref. 27 |
| $L_c$ | Nematic length | 20 − 100 μm | Ref. 27 & 28 |
| $T_0$ | Traction coefficient | 0.1 − 1 kPa | Ref. 27 & 28 |
| $\eta$ | Cell viscosity | 0.6 − 60 MPa·s | Ref. 27 & 28 |
| $\xi$ | Friction coefficient | 150 − 1200 Pa·s/μm | Ref. 27 & 28 |
| $-\zeta_0$ | Active contractility | ≤ 50 kPa | Ref. 27 & 28 |
| $\Gamma_0$ | Surface tension in contractile ring | ≤ 15 mN/m | Ref. 31 |
| Dimensionless | | | |
| $\alpha$ | Friction-to-viscosity ratio | 0 − 2 | |
| $\beta$ | Nematic-length-to-thickness ratio | 4 − 20 | |
| $-\zeta$ | Dimensionless active | ≤ 500 | |

| | contractility | |
|---|---|---|
| $\gamma_0$ | Dimensionless surface tension | $\leq 7.5$ |

## 4. Results and discussions

In this section, we analyze the spreading velocity of the epithelial monolayer and apply the results to the study of the re-epithelialization process based on the analytical solutions derived in the last section. First, we investigated the signs of velocity in the radial direction to determine whether the epithelial sheet approaches the center of the circular gap. For the first set of solutions derived by $\alpha = 0$, we plot the contour of $v_r(R_{in}, \zeta, \gamma) = 0$ with respect to the cases that $\gamma = \gamma_0$, $\gamma = \gamma_1|\kappa|$, and $\gamma = \gamma_2\kappa^2$. In Fig. 2 (a), (b), and (c), each contour stands for a specific value of $\gamma_m$ ($m = 0, 1, 2$). Each contour of $v_r(R_{in}, \zeta, \gamma) = 0$ divides the $R_{in} - \zeta$ plane into two parts: the left part stands for the case where $v_r < 0$, indicating that the epithelial monolayer migrates towards the gap center; while the right part stands for the case where $v_r > 0$, meaning that the epithelial cells are migrating away from the gap center. As the values of $\gamma_m$ and $m$ increase, the area of the region where $v_r < 0$ also increases, which indicates that the purse-string contraction is one of the main driving forces of the gap closure. For a circular gap with a specific size of $R_{in}$, the monolayer front cannot migrate towards the gap center ($v_r > 0$) as the tissue active contractility $-\zeta$ reaches a critical value $-\zeta_{cr}$. This implies that overly high active contractility is unfavorable for gap closure. The critical cell contractility $-\zeta_{cr}$ is a linear function of the coefficient of the purse-string contraction, as plotted in Fig. 3, with a slope of $-\zeta_{cr}/\gamma_m$ equal to $2/R_{in}^{m+1}\{1 - [\frac{K_0(R_{in})}{K_1(R_{in})}]^2\}$. For an epithelial monolayer with fixed cell contractility, there exists a critical gap radius $R_{cr}(\gamma)$, beyond which migration towards the gap center is unfavorable. The critical radius $R_{cr}(\gamma)$ can be derived by looking for the smaller solution to the equation $v_r = 0$ with respect to certain values of $\zeta$, which is plotted in Fig. 4. From Fig. 4, it can be concluded that higher levels of purse-sting contraction can lead to larger values of $R_{cr}$, while higher active cell contractility leads to smaller $R_{cr}$.

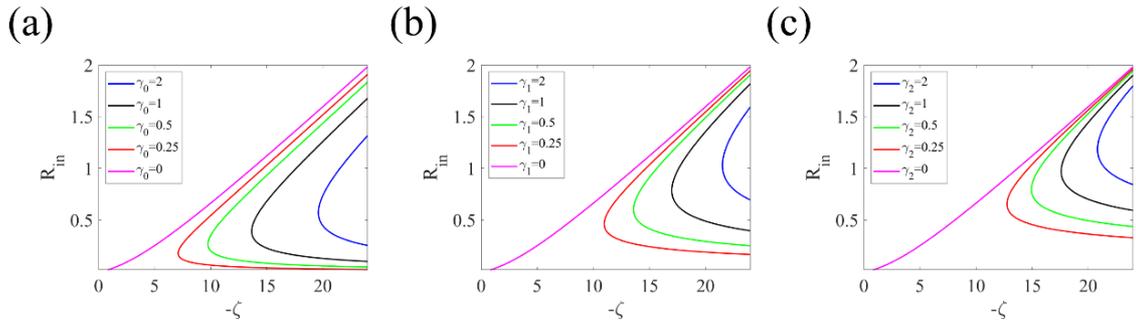

**Fig. 2.** The contours of $v_r(R_{in}, \zeta, \gamma) = 0$ in the $R_{in} - \zeta$ plane under the conditions of (a) the purse-string contraction $\gamma = \gamma_0$, (b) the purse-string contraction $\gamma = \gamma_1|\kappa|$, and (c) the purse-string contraction $\gamma = \gamma_2\kappa^2$. Each contour stands for a specific value of $\gamma_m$ ($m = 0, 1, 2$). The other parameters used in this plot are $\alpha = 0$ and

$\beta = 5$.

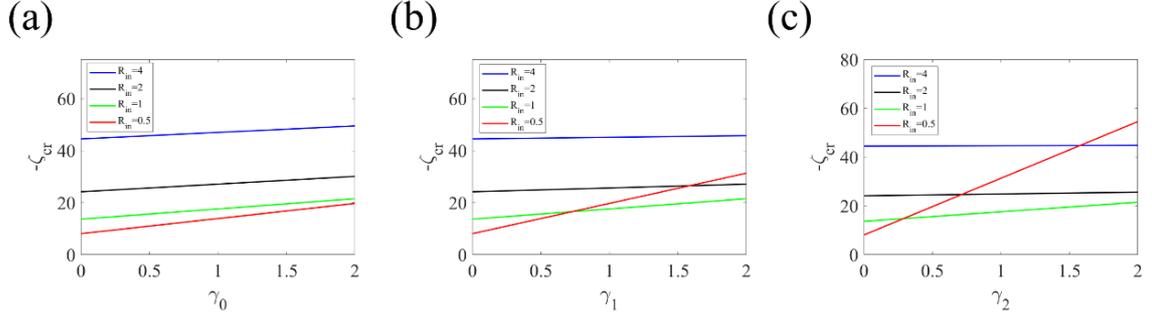

**Fig. 3.** The critical value of the intercellular contractility $-\zeta_{cr}$ (as a function of purse-string contraction coefficient $\gamma_m$ that leads to $v_r(R_{in}, \zeta, \gamma) = 0$ under the conditions of (a) purse-string contraction $\gamma = \gamma_0$, (b) purse-string contraction $\gamma = \gamma_1|\kappa|$, and (c) purse-string contraction $\gamma = \gamma_2\kappa^2$. Each curve stands for a specific value of $\gamma_m$ ($m = 0, 1, 2$). The other parameters used in this plot are $\alpha = 0$, $R_{in} = 1$, and $\beta = 5$.

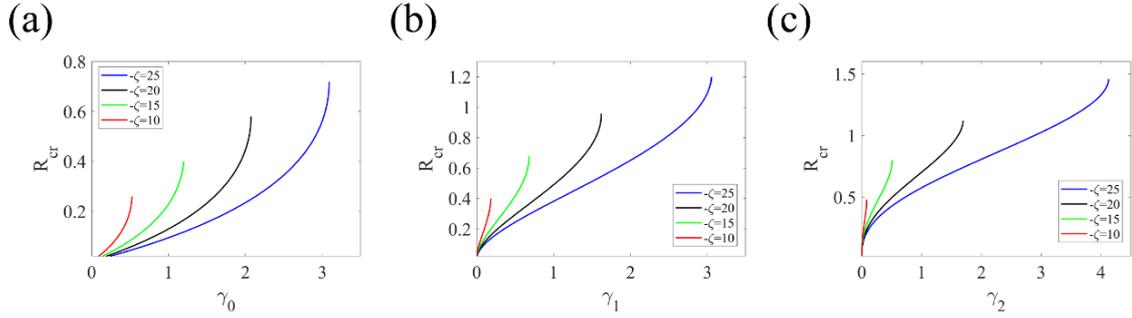

**Fig. 4.** The critical value of the gap radius $R_{cr}$ (as a function of $\gamma_m$) that leads to $v_r(R_{in}, \zeta, \gamma) = 0$ under the conditions of (a) the purse-string contraction $\gamma = \gamma_0$, (b) the purse-string contraction $\gamma = \gamma_1|\kappa|$, and (c) the purse-string contraction $\gamma = \gamma_2\kappa^2$. Each curve stands for a specific value of $\gamma_m$ ($m = 0, 1, 2$). The other parameters used in this plot are $\alpha = 0$ and $\beta = 5$.

For the second set of solutions derived by $\alpha = 2$, one can arrive at the following conclusion by exploiting Eq. (40):

$$v_r = \frac{\zeta\{2K_0(R_{in})K_1(R_{in})K_0(2R_{in}) - K_1(2R_{in})[K_0^2(R_{in}) + K_1^2(R_{in})]\}}{4K_1'(2R_{in})K_1^2(R_{in})} + \frac{\gamma(R_{in})K_1(2R_{in})}{2R_{in}K_1'(2R_{in})} <$$

$$-\frac{\zeta K_1(2R_{in})[K_1(R_{in}) - K_0(R_{in})]^2}{4K_1'(2R_{in})K_1^2(R_{in})} < 0, \tag{62}$$

which implies that the epithelial monolayer unconditionally migrates towards the gap center for this case.

Next, we applied these derived analytical solutions to investigate the kinetics of the gap closure of the epithelial monolayer. We have the following relation between the gap radius and the radial component of the velocity vector:

$$\frac{dR_{in}}{dt} = v_r(R_{in}). \tag{63}$$

By integrating the equation, we have:

$$\int_{R_{in}^0}^{R_{in}^1} \frac{dR_{in}}{v_r(R_{in})} = \int_0^t dt. \tag{64}$$

The above equation can be used to describe the gap closure process as a function of time; the results are plotted in Fig. 5 and Fig. 6. The nondimensionalized gap radius is plotted as a function of dimensionless time with respect to different forms of the purse-string contraction $\gamma = \gamma_m |\kappa|^m$ in Fig. 5 for $\alpha = 0$ and in Fig. 6 for $\alpha = 2$. By observing these figures, one can see that the power law of the purse-string contraction has a great impact on the gap closure kinetics. On the one hand, as the value of the power $m$ increases, the time spent on the gap closure decreases no matter whether the viscosity dominates over the friction. On the other hand, the power law dramatically steepens the radius-time curve when the gap is nearly closed. This steepened shape of the curves corresponds to the perimeter-time curve recorded in previous experiments[38].

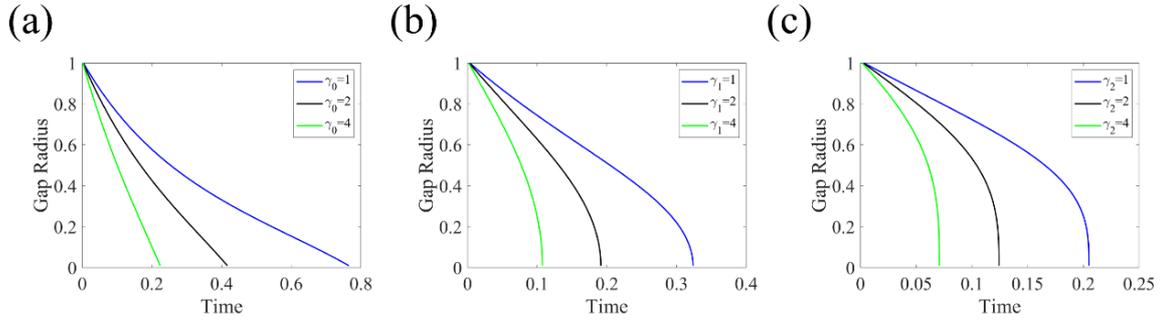

**Fig. 5.** The nondimensionalized gap radius as a function of dimensionless time for $\alpha = 0$ with respect to the conditions of (a) the purse-string contraction $\gamma = \gamma_0$, (b) the purse-string contraction $\gamma = \gamma_1 |\kappa|$, and (c) the purse-string contraction $\gamma = \gamma_2 \kappa^2$. Each curve stands for a specific value of $\gamma_m$ ($m = 0, 1, 2$). The other parameter used in this plot is $\beta = 5$, and $-\zeta = 6$.

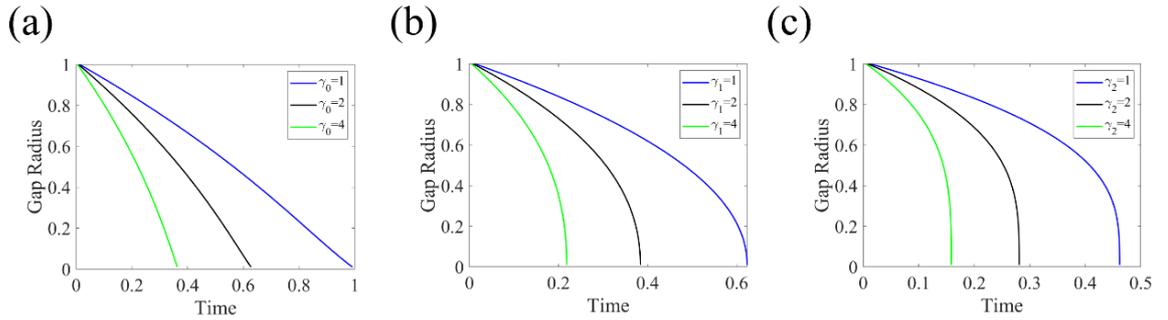

**Fig. 6.** The nondimensionalized gap radius as a function of dimensionless time for $\alpha = 2$ with respect to the conditions of (a) the purse-string contraction $\gamma = \gamma_0$, (b) the purse-string contraction $\gamma = \gamma_1 |\kappa|$, and (c) the purse-string contraction $\gamma = \gamma_2 \kappa^2$. Each curve stands for a specific value of $\gamma_m$ ($m = 0, 1, 2$). The other parameter used in this plot is $\beta = 20$, and $-\zeta = 6$.

Since tissue morphological instabilities are almost inevitable in collective cell motion,

the loss of the circularity of the gap boundary is also of interest. By solving Eq. (54) ~ (56) and (60) for $R_{in}^0 = 1$, we obtain the perturbation in the spreading velocity relative to that in the gap radius $\delta \tilde{v}_n / \delta \tilde{R}_n$ at the monolayer front for different wavenumbers. Here, the purse-string contraction is chosen to have the form $\gamma = \gamma_2 \kappa^2$ according to the measurements from previous experiments[33]. To visualize such perturbations, we plot $\mathbf{r} = \mathbf{r}_0 + \delta \mathbf{v} dt$ at $\mathbf{r}_0 = 1, 1.05,$ and $1.1$ in Fig. 7. The perturbated velocity shows a $360°/n$ rotational symmetry as one could expect. The morphologies of the perturbated configuration $\mathbf{r}$ are quite similar for the cases of $\alpha = 0$ and $\alpha = 2$, despite the perturbated morphologies for $\alpha = 2$ being slightly more rounded than those for $\alpha = 0$, suggesting $\delta \tilde{v}_n / \delta \tilde{R}_n$ having larger magnitudes for the case of $\alpha = 0$ than $\alpha = 2$.

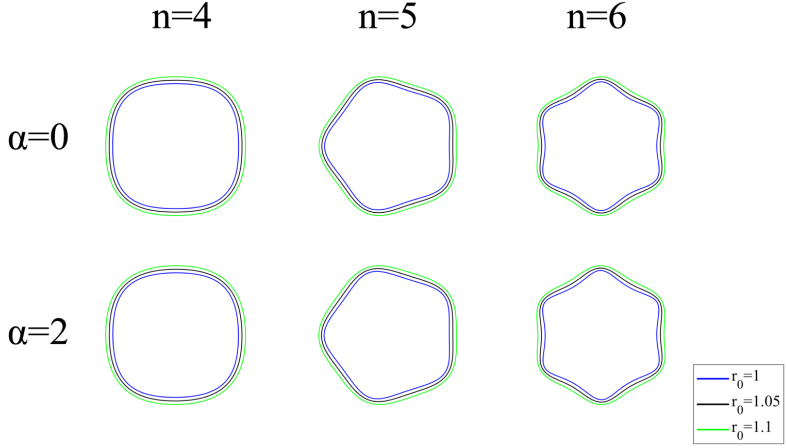

**Fig. 7.** The perturbation in the morphology of the circular front of the migrating epithelial layers. We have plotted the perturbated shapes at $r_0 = 1, 1.05, \& 1.1$ in the original configuration. The cases of the wavenumber $n$ equaling $2, 4, \& 6$ are calculated with both cases of $\alpha = 0$ and $\alpha = 2$. The ring contraction is $\gamma = \gamma_2 \kappa^2$, and $-\zeta = 6$. The other parameter used in this plot is $\beta = 5$ for $\alpha = 0$ and $\beta = 20$ for $\alpha = 2$.

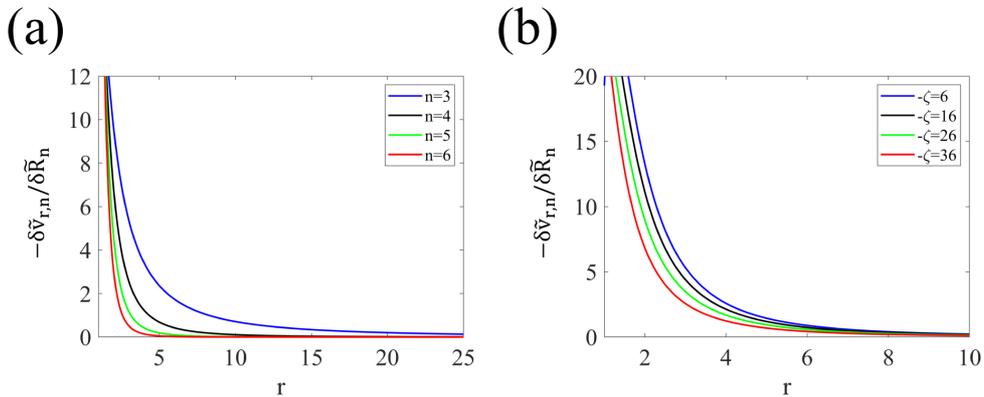

**Fig. 8.** For the case of $\alpha = 0$, $-\delta \tilde{v}_{r,n} / \delta \tilde{R}_n$ is a function on the radial coordinate $r$, with respect to (a) different wavenumbers, the other parameters are $\beta = 5$, $-\zeta = 6$, $\gamma_2 = 1$, and $R_{in}^0 = 1$; and (b) different intercellular contractility, the other parameters are $n = 4$, $\beta = 5$, $\gamma_2 = 1$, and $R_{in}^0 = 1$.

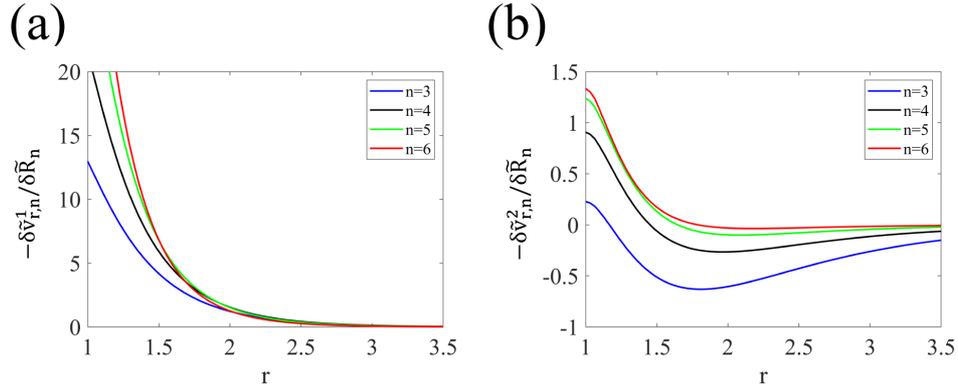

**Fig. 9.** For the case of $\alpha = 2$, $-\delta\tilde{v}_{r,n}^{1,2}/\delta\tilde{R}_n$ is a function of the radial coordinate $r$ with respect to different wavenumbers: (a) $-\delta\tilde{v}_{r,n}^1/\delta\tilde{R}_n$ and (b) $-\delta\tilde{v}_{r,n}^2/\delta\tilde{R}_n$. The other parameters are $\beta = 20$, $-\zeta = 6$, $\gamma_2 = 1$, and $R_{in}^0 = 1$.

Next, we examine the case of $\alpha = 0$ and examine the perturbation in the spreading velocity in the radial direction $\delta\tilde{v}_{r,n}/\delta\tilde{R}_n$ as a function of the radial coordinate $r$ with respect to different values of the wavenumber $n$ and the intercellular contractility $-\zeta$. From Fig. 8, we can see that the magnitudes of $-\delta\tilde{v}_{r,n}/\delta\tilde{R}_n$ decrease monotonically and eventually decay to zero as the value of $r$ increases. The increasing value of $n$ and decreasing value of $-\zeta$ both lead to a faster decay in $-\delta\tilde{v}_{r,n}/\delta\tilde{R}_n$. For the case of $\alpha = 2$, we investigate $-\delta\tilde{v}_{r,n}^{1,2}/\delta\tilde{R}_n$ (in Eq. (61d)) as a function of $r$ with respect to different wavenumbers. By looking into Fig. 9 (a) and (b), one can see that the magnitudes of $-\delta\tilde{v}_{r,n}^1/\delta\tilde{R}_n$ are much larger than those of $-\delta\tilde{v}_{r,n}^2/\delta\tilde{R}_n$ for all the wavenumbers $n$ we study, which suggests $\delta\tilde{v}_{r,n}^1$ play a much more important role than $\delta\tilde{v}_{r,n}^2$. Since $\delta\tilde{v}_{r,n}^1$ corresponds to the cosine series according to Eq. (61d) explains why the morphologies of the perturbated circular front for $\alpha = 2$ are quite similar to those obtained by letting $\alpha = 0$. Moreover, the perturbations derived for $\alpha = 2$ decay much faster than the case of $\alpha = 0$ for wavenumbers equal to 3 and 4. However, for the wavenumber $n$ equal to 6, the decay rates for both cases are quite the same: the magnitudes of $\delta\tilde{v}_{r,n}/\delta\tilde{R}_n$ both reach roughly zero at $r = 3.5$.

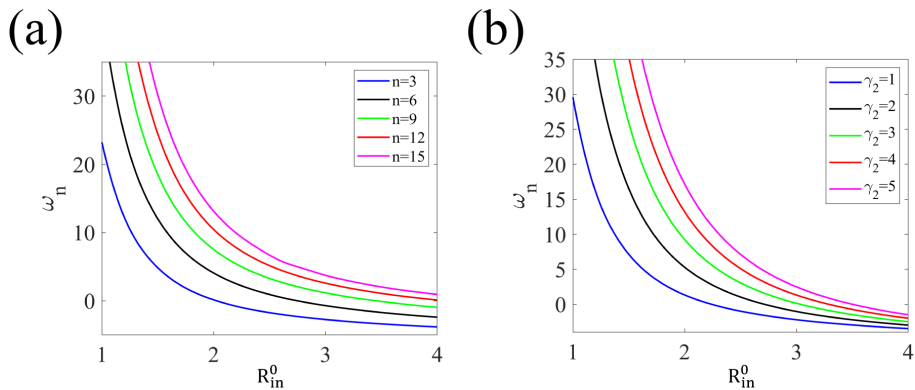

**Fig. 10.** For the case of $\alpha = 0$, the growth rate $\omega_n$ is a function of the gap size $R_{in}^0$ with respect to (a) different values of $n$, with other parameters being chosen as $\beta = 5$, $-\zeta = 6$, and $\gamma_2 = 1$; (b) different values of $\gamma_2$, with other parameters being chosen as $\beta = 5$, $-\zeta = 6$, and $n = 4$.

Finally, we define the $n$-th order growth rate $\omega_n$ of the tissue shape perturbation for the case of $\alpha = 0$:

$$\omega_n = -\left[\frac{\partial v_r^0(R_{in}^0)}{\partial r} + \frac{\delta \tilde{v}_{r,n}(R_{in}^0)}{\delta \tilde{R}_n}\right]. \tag{65}$$

It can be observed from Eq. (65) that the growth rate explicitly depends on the wavenumber $n$ and the gap radius $R_{in}^0$. Thus, we plot the growth rate as a function of $R_{in}^0$ with respect to an ascending sequence of wavenumbers. Fig. 10 (a) shows that the growth rates are mostly positive for the chosen values of $R_{in}^0$, meaning that $\delta \tilde{R}_n$ usually has the same sign with $d\delta \tilde{R}_n/dt$. Additionally, as the gap radius $R_{in}^0$ increases, the growth rates tend to converge for all the wavenumbers we study. Moreover, larger wavenumbers lead to more rapid changes in the growth rate as $R_{in}^0$ varies. This indicates that the perturbations with larger wavenumbers grow faster than perturbations with smaller wavenumbers. We also plot the growth rate with respect to different values of the purse-string contraction in Fig. 10 (b), from which one could tell that the stronger ring contraction is in favor of the faster perturbation growth rate.

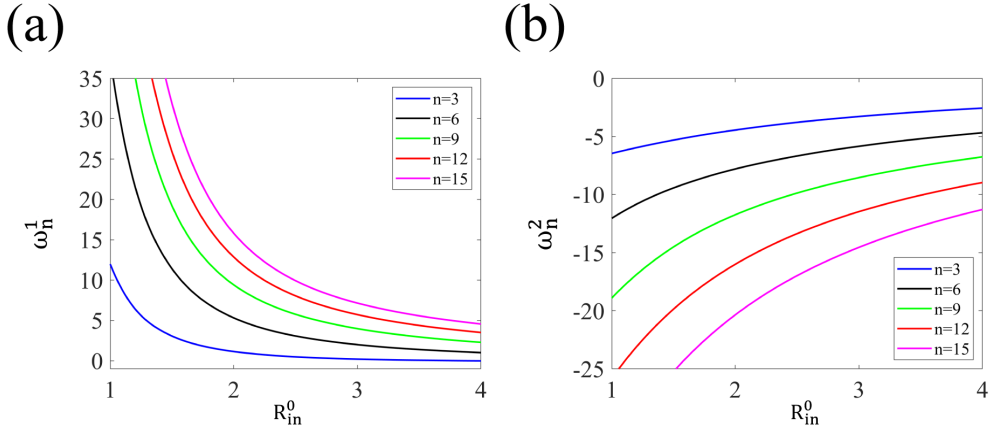

**Fig. 11.** For the case of $\alpha = 2$, (a) the growth rate $\omega_n^1$ and (b) the growth rate $\omega_n^2$ are both functions of the gap size $R_{in}^0$, and the parameters are chosen as $\beta = 20$, $-\zeta = 6$, and $\gamma_2 = 1$.

For the case of $\alpha = 2$, the $n$-th order growth rate $\omega_n$ of the tissue shape perturbation has two components that correspond to the cosine and sine series in Eq. (61):

$$\omega_n^1 = -\left[\frac{\partial v_r^0(R_{in}^0)}{\partial r} + \frac{\delta \tilde{v}_{r,n}^1}{\delta \tilde{R}_n}\right], \tag{66a}$$

$$\omega_n^2 = -\left[n\frac{\partial v_\theta^0(R_{in}^0)}{R_{in}^0} + \frac{\delta \tilde{v}_{r,n}^2}{\delta \tilde{R}_n}\right], \tag{66b}$$

which are plotted in Fig. 11 as functions of $R_{in}^0$ with respect to different wavenumbers. By looking into Fig. 11 (a), we can see a pattern that is similar to the case of $\alpha = 0$ plotted in Fig. 10 (a). What is different is that the growth rates $\omega_n^1$ in the case of $\alpha = 2$ are always positive, yet for the case of $\alpha = 0$, the growth rates $\omega_n$ can possess negative values. It implies that for the case of $\alpha = 2$, if $\delta \tilde{R}_n < 0$, the perturbation always leads the monolayer front to spread towards the gap center; while for or the case of $\alpha = 0$, the perturbation could steer the monolayer front to migrate away from the gap center under certain conditions as shown in Fig. 10. For the growth rate $\omega_n^2$, its value is dominated by the term $-n \frac{\partial v_\theta^0(R_{in}^0)}{R_{in}^0}$ since $\frac{\delta \tilde{v}_{r,n}^2}{\delta \tilde{R}_n}$ is relatively small (shown in Fig. 12 (b)); hence, the value of $\omega_n^2$ monotonically decreases as the wavenumber $n$ increases.

## 5. Conclusions

In this article, we study the migration of an epithelial monolayer with a circular spreading front by an active polar fluid model and apply the results to study the gap closure process. By studying the axisymmetric case, we obtain two sets of analytical solutions to the velocity and polarity fields under two special cases: one is the case that the cellular viscosity dominates over the cell-substrate friction, which corresponds to the dimensionless model parameter $\alpha$ equal to zero; the other is that the friction cannot be neglected and is in certain relation with the cellular viscosity, which corresponds to $\alpha = 2$. For the case that viscosity dominates over friction, there exists a critical gap radius and active contractility for the initiation of the gap closure process. Moreover, the power law of the purse-string contraction has a great impact on the gap-closure kinetics. By choosing the right power of the curvature-dependence in the purse-string contraction, one can replicate the gap radius-time curve recorded from past experiments. Finally, we discuss the loss of circularity of the spreading front during the gap closure process by solving the perturbated equations for the polarity and velocity fields. For $\alpha = 0$, we have found that large wavenumbers and intercellular contractility lead to large perturbations in the spreading velocity. We have analyzed how the growth rate of the tissue shape perturbation, as a function of gap radius, is affected by the wavenumber and actin-ring contraction. For $\alpha = 2$, we have studied how both growth rates change with respect to different gap sizes and wavenumbers. Although we neglect the rich mechano-biochemical process occurring during the whole re-epithelialization process, our analytical solutions still capture some experimental observations from previous studies.

## 6. Declaration of competing interest

The corresponding author declares that there is no competing financial interest in the work reported in this paper.

## 7. Acknowledgement

T. Z. acknowledges the funding support from the Southern University of Science and Technology, China (sustech.edu.cn).

## 8. Declarations

**Ethical Approval** Not applicable.
**Author contributions** T.Z. conceptualized the research; T.Z. and H.Y. performed the research; T.Z. and H.Y. wrote the manuscript.
**Funding** Not applicable.
**Availability of data and materials** Not applicable.